\documentclass[conference]{IEEEtran}
\usepackage{graphicx}
\usepackage{caption}
\usepackage{subcaption}
\usepackage{amsmath}
\usepackage{soul,color}
\usepackage{todonotes}
\usepackage{url}
\usepackage{mathtools}
\usepackage{calc}
\usepackage{xparse}
\usepackage{algorithm}
\usepackage{algpseudocode}
\usepackage{pifont}
\usepackage{makeidx} 

\begin{document}
\title{Dynamic Base Station Repositioning to Improve Performance of Drone Small Cells}

\author{\IEEEauthorblockN{Azade Fotouhi\IEEEauthorrefmark{1}\IEEEauthorrefmark{2},
Ming Ding\IEEEauthorrefmark{2} and
Mahbub Hassan\IEEEauthorrefmark{1}\IEEEauthorrefmark{2}}
\IEEEauthorblockA{\IEEEauthorrefmark{1}School of Computer Science and Engineering\\
University of New South Wales (UNSW), Sydney, Australia\\ 
Email: \{a.fotouhi, mahbub.hassan\}@unsw.edu.au}
\IEEEauthorblockA{\IEEEauthorrefmark{2}Data61, Australia\\
Email: \{azade.fotouhi, ming.ding, mahbub.hassan\}@data61.csiro.au}
}

\maketitle

\begin{abstract}
Thanks to the recent advancements in drone technology, it has become viable and cost-effective to quickly deploy small cells in areas of urgent needs by using a drone as a cellular base station. In this paper, we explore the benefit of dynamically repositioning the drone base station in the air to reduce the distance between the BS and the mobile user equipment, thereby improving the spectral efficiency of the small cell. In particular, we propose algorithms to autonomously control the repositioning of the drone in response to users activities and movements. We demonstrate that, compared to a drone hovering at a fixed location, dynamic repositioning of the drone moving with high speed can increase spectral efficiency by 15\%. However, considering the tradeoff between spectral efficiency and energy efficiency of the drone, we show that 10.5\% spectral efficiency gain can be obtained without negatively affecting energy consumption of the drone.  
\end{abstract}

\IEEEpeerreviewmaketitle

\section{Introduction}

A drone is an unmanned aerial vehicle designed to be flown either through remote control or autonomously using embedded software and sensors, such as GPS. Historically, drones had been used mainly in military for reconnaissance purposes, but with recent developments in light-weight drones operated with batteries, many civilian applications are emerging. Use of drones to deploy small cells in areas of urgent needs is one of the most interesting applications currently being studied by many researchers \cite{Namuduri,6399121,7122576,yaliniz2016efficient}. The greatest advantage of this approach is that drones can be fitted with small cell base station (BS) equipment and sent to a specific target location immediately without having to deploy any infrastructure. 

Recent studies on drone small cells mainly focussed on finding the optimum location in the air for the drone to hover while serving the target area with a given population and traffic demand. For example, authors in \cite{mozaffari2015drone} studied the problem of finding the optimal height for a drone to cover a fixed target area with the minimum transmission power. Sharma et al. {\cite{7451189}} considered a more complex problem involving optimal deployment of multiple drones between the macro and small cell tiers for improving the coverage and capacity of the whole system. Yet, other researchers attempted to optimize the locations in the air for drones to relay traffic between BS and user equipment (UE) to maximize the overall data rates {\cite{7122575}}. In all of these studies, only the hovering position in the air is optimized without considering any dynamic repositioning \textit{during} the service.

The goal of this study is to explore the benefit of dynamic repositioning of the drone during the service in response to the dynamic users activities and mobility. The key idea is to exploit the flying capability and agility of light-weight drones for making the BS continuously `chasing' the current location of active users within the cell, thereby reducing the distance between BS and UE~\footnote{Note that the concept of the BS chasing the active UEs is not possible with existing infrastructure-based cellular networks, which makes this particularly interesting and appealing to pursue for drone small cells.}. By bringing the BS closer to the UE, we can not only reduce the signal attenuation, but also increase the probability of line-of-sight (LoS) for a given altitude of the drone. The combination of these two effects is expected to increase the data rate and hence spectral efficiency of the drone cell. 

It is obvious that, for the proposed dynamic repositioning of the drone, autonomous control with embedded software is preferred over ground-based remote control. This means, we need to design algorithms for the drone to continuously reposition itself within the cell in response to user activity and mobility. An associated challenge is to ensure that drone energy consumption is not significantly increased due to dynamic repositioning compared to the strict hovering at the same location as considered in previous studies. To this end, we propose multiple algorithms and evaluate their performances using simulations. We show that the proposed algorithms can increase the spectral efficiency of drone small cells by 15\% while being more energy efficient than a fixed one in terms of communication energy efficiency. Without negatively affecting the drone mechanical energy consumption, the spectral efficiency can be improved by 10.5\% using our proposed algorithms.
    
Our contributions can be summarized as follows:
\begin{itemize}
\item We introduce and explore the concept of \textit{dynamic repositioning} as a new method for improving spectral efficiency of drone small cells. 
\item We propose and evaluate three practically realizable algorithms for autonomous dynamic repositioning of drone BSs. We show that, for all of these algorithms, gain in spectral efficiency increases as a function of drone speed and as such there exists a tradeoff between spectral efficiency and energy efficiency of the drone.
\item We show that, dynamic repositioning can deliver 10.5\% more spectral efficiency without any negative impact on energy efficiency.

\end{itemize}
 

\section{System Model} \label{sec:systemmodel}

Let us consider a square area with a width of $w$ (in meter) to be covered by a drone flying at a height of $h$ (in meter) with a constant speed of $v$ (in m/s). Here, we assume the height and the speed of drone are fixed, and it can only change its direction of moving. The drone is responsible to provide high data-rate services for the users in the square area. 

In this considered area, there are $U$ users randomly moving according to the Random Way Point model (RWP). The download traffic model for each user follows a traffic model recommended by 3GPP \cite{3gpp36814}. The interval between each user's request follows Poisson Point distribution with a mean value of $\lambda$ (in sec). The request arrival time of a user $u$ is denoted by $R_u= \{ t_{u,1},t_{u,2},t_{u,3},\dots \} $, where the interval between two consecutive arrival times is modeled as the Poisson Point distribution mentioned above.
 The data size of each request is denoted by $s$ (in MBytes). During the time period from request time of a user to the end of download time, the user is called an \textit{active} user. We assume that users' position information is piggybacked on control signals.
 
We further assume that the drone is transmitting data to users with a transmission power of $P_{tx}$ (in watt), $f$ (in Hz) denotes the working frequency of the drone and $B$ (in MHz) is the total bandwidth available for the drone communication.

Moreover, we denote the number of active requests at time $t$ and the amount of allocated bandwidth to the $i$-th request by $q_t$ and $b_i$($1\leq i \leq q_t$, $0\leq b_i \leq B$ ), respectively.

The propagation channel is modeled by probabilistic LoS, in which the probability of having a LoS connection between a user and the drone is as follows \cite{al2014optimal}
\begin{equation}
P^{LoS}(h,r) = \frac{1}{1+\alpha exp(-\beta\theta -\alpha])}
\label{eq:plos}
\end{equation}
where $\alpha$ and $\beta$ depend on the environment, $\theta$ equals to $arctan(h/r)$ in degree, $h$ denotes the drone height and $r$ measures the distance between the ground user and the projection of drone location onto the ground. Clearly, the probability of having a NLoS (Non Line of Sight) connection is $P^{NLoS} (h,r)= 1 - P^{LoS}(h,r)$. 

The path loss in dB is then modeled as 
\begin{equation}
\eta_{path}(d)= A+ 10\gamma\log_{10}(d)
\label{eq:pathloss}
\end{equation}  
where the string variable \textit{"path"} takes the value of “LoS” and “NLoS” for the LoS and the NLoS cases, respectively. In addition, $A$ is the path loss at the reference distance (1 meter) and $\gamma$ is the path loss exponent, both obtainable from field tests \cite{3gpp36828}. Besides, $d$ is the distance (in meter) between the transmitter and the receiver ($d = \sqrt{h^2+r^2}$), $A$ and $\gamma$ take different values for the LoS and the NLoS cases.

The \textit{spectral efficiency} (bps/Hz) of an active user $i$ at distance $d$ from the drone can be formulated according to the Shannon Capacity Theorem as
\begin{align}
 \begin{split}
\varphi_i(d) = \log_2 (1+\frac{S_{i}^{path}(d)}{N_i}) 
\end{split}
 \label{eq:individualspec}
\end{align}
where $N_i$ (in watt) represents the total noise power including the thermal noise power and the user equipment noise figure, which is given by \cite{thermalnoise} 
\begin{equation}
N_{i} = 10^{ \frac{-174+\delta_{ue}}{10}}.(b_i).10^{-3}
\label{eq:noise}
\end{equation}
where $\delta_{ue}$ is the user equipment noise figure (in dB), and $b_i$ is the user's allocated bandwidth.

In (\ref{eq:individualspec}), $S_{i}^{path}(d)$ (in watt) indicates the received power of the $i$-th user from the drone, which can be obtained by
\begin{align}                                                                                                                                          
S_{i}^{path}(d) =\frac{b_i}{B} (P_{tx}.10^{\frac{-\eta_{path}(d)}{10}})
\label{eq:rcvpower}
\end{align}   

Plugging (\ref{eq:rcvpower}) and (\ref{eq:pathloss}) into (\ref{eq:individualspec}), we can simplify the average spectral efficiency as
\begin{align}
 \begin{split}
 \bar{\varphi}_i(d) =&  P^{LoS}(h,r).\log_2 \Big(1+\frac{S_{i}^{LoS}(d)}{N_{i}}\Big)\\
 +&P^{NLoS}(h,r).\log_2 \Big(1+\frac{S_{i}^{NLoS}(d)}{N_{i}}\Big)\\
\end{split}
 \label{eq:simpleindividualspec}
\end{align}

\begin{figure}[!t]
\centering
  \includegraphics[scale=0.40]{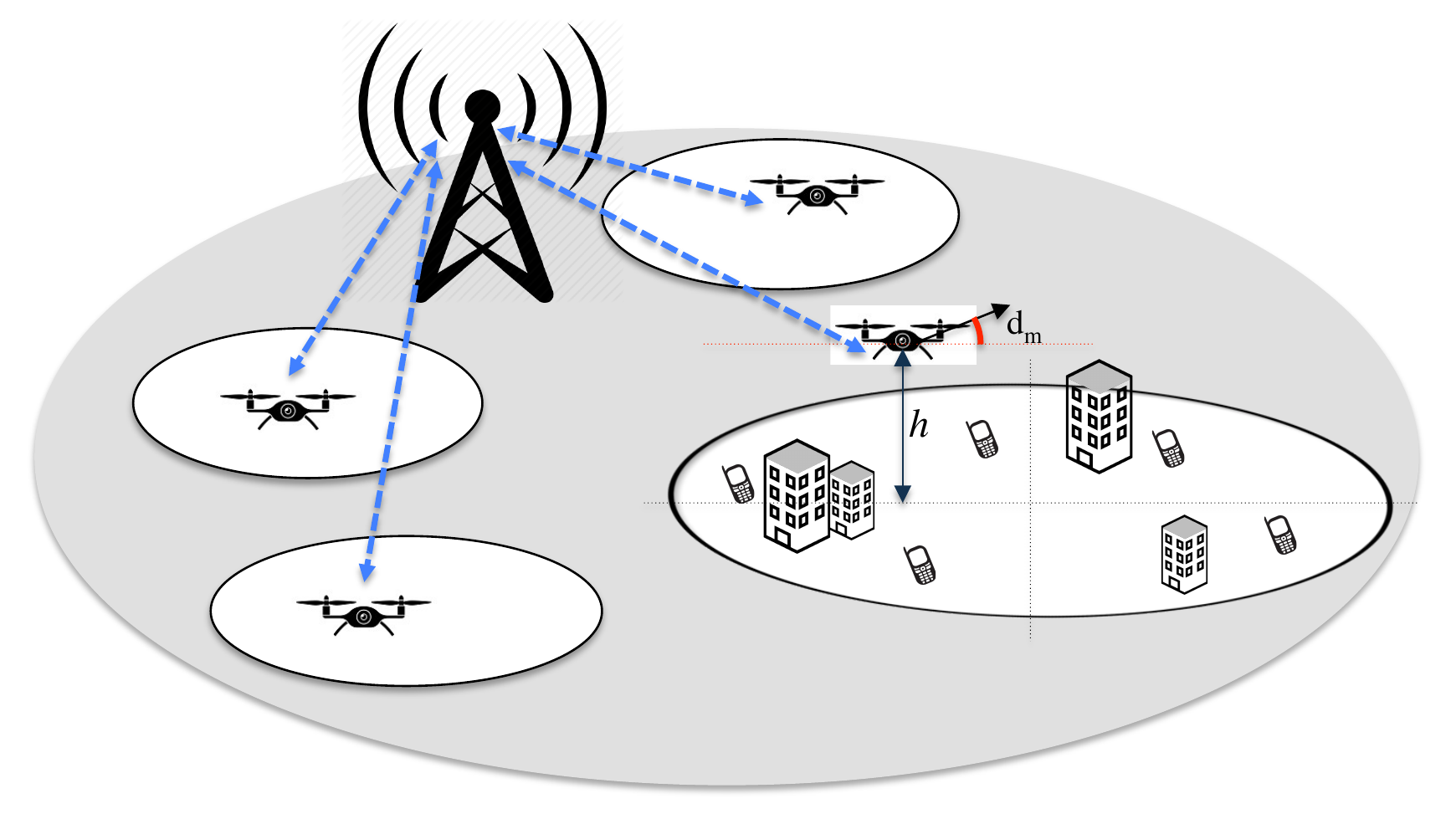}
\caption{Finding the appropriate movement direction by the drone BS}
\label{fig:angle}
\end{figure}

The drone updates the location of itself continuously in order to maximize the spectral efficiency of the system when it is transmitting data to the active users. We divide the operation time into timeslots of $\Delta t$ seconds. At the start of each timeslot, the drone finds a direction to move forward during the timeslot. The direction is selected according to the location of active users to maximize the spectral efficiency of the cell
(See figure {\ref{fig:angle}}). As mentioned before, one important challenge is to find the directions and plan the route for the drone considering the spectral efficiency and the energy consumption. Our proposed methods are presented in the following section.

\section{The Proposed Algorithms}\label{sec:movingstg}


We propose three different movement strategies to improve the spectral efficiency of the system while taking the energy consumption of the drone into account. These methods will be presented in the following subsections.
\subsection{Equal Bandwidth Division}

In this strategy, the drone serves all active users with equal shares of resources. At each defined time slot $\Delta t$, the drone simply divides the total bandwidth equally among all active users and chooses the movement direction to maximize the achievable spectral efficiency of the system. 

The direction is defined based on the angle created between the movement line of drone and the horizontal line parallel with x-axis. The drone is able to choose any direction in $[ 0 \ 2\pi)$, however in order to reduce the complexity of the problem, an angle step is defined and denoted by $\Delta g$. For instance, if the angle step is $\Delta g = \pi/M$, then $2M$ candidate directions, $\{0,\frac{\pi}{M}, \frac{2\pi}{M} , \dots , 2\pi - \frac{\pi}{M}\}$, would be examined by the drone. 
At first, the drone finds its candidate locations following movement towards each possible direction. Then it calculates the new average spectral efficiency for all users at each candidate location and finds the direction/candidate location that maximizes the obtainable average spectral efficiency. Moreover, the average spectral efficiency for the current location of the drone is calculated as well. If the average spectral efficiency of the current location is larger than the achievable value of the selected candidate location, then the drone stays hovering at its current location. Otherwise, it starts moving towards the new direction. The drone specifies a new direction for the next time slot in a similar way. When transmission for a request finishes, the amount of released bandwidth is shared equally between remaining requests. If there is no active user the drone stays hovering at its current location. 


\subsection{Nearest User First}
In this strategy, instead of allocating equal bandwidth to all active users, just one user receives the whole bandwidth at each time slot. It can be easily shown that the allocation of whole bandwidth to the active user having the highest spectral efficiency will maximize the spectral efficiency at each time slot. Among all active users, the nearest user to the current location of drone BS has the highest value of spectral efficiency. Following this, the drone allocates the whole bandwidth to the nearest active user and moves towards it during each time slot. If more than one user have the same distance from the drone BS, the one with the earliest request time is selected. The drone transmits data to the selected user and repeats the decision making process at each time slot. When there is no request to serve, the drone stays hovering at its current location.

\subsection{Least Buffer First}
The above two strategies focus on the spectral efficiency and try to increase it by means of the drone movement. In this strategy, we consider another performance metric, i.e., data buffer, to control the drone's movement. At each time slot, the drone finds the user with the least remaining data to send, allocates the whole bandwidth to it, and moves towards the user. If there are several users with the same remaining buffer, the one with the highest spectral efficiency is chosen.

Considering the same data size for all requests, this strategy is actually the same as the FIFO (First In First Out) strategy. Therefore, this strategy tries to reduce the latency for all users.

 \section{Performance Metrics} \label{sec:perfmet}
In this section, different performance metrics are defined to evaluate our proposed strategies. These performance metrics are used to compare our movement strategies for the mobile drone BS with a drone BS which is hovering above the center of target area.
 
 \subsection{Spectral Efficiency Gain}
One important goal of the mobile drone is to provide high spectral efficiency for the area of interest. \textit{Spectral Efficiency Gain (SEG)} represents the percentage of spectral efficiency improvement of the mobile drone BS compared with the fixed one, defined as follows
 
 \begin{align}
 \begin{split}
SEG= \frac{100*(\bar{\varphi}_{m} - \bar{\varphi}_{s})}{ \bar{\varphi}_{s}}
\end{split}
 \label{eq:gain}
 \end{align}
where $\bar{\varphi}_{m}$ and $ \bar{\varphi}_{s}$ denote the average spectral efficiency value of the mobile drone BS and the fixed one, respectively.
 
 \subsection{Energy Consumption Metrics}\label{sec:engmodel}
Because of the importance of energy consumption for drones, we consider the energy efficiency issue of our proposed strategies. There are two main sources of energy consumption for the drone BS i.e., communication energy consumption and mechanical energy consumption. The communication energy depends on the transmission power and the communication time, which can be expressed by
\begin{align}
 \begin{split}
E_{comm} = P_{tx}.t_{communication}
\end{split}
 \label{eq:comeng}
 \end{align}
 
Furthermore, the \textit{Communication Energy Efficiency (CEE)} (in bits/joule) is defined to be the ratio of the number of total transmitted bits over the total communication energy, i.e.,
\begin{align}
 \begin{split}
CEE = \frac{(\sum_{u=1}^{U}{|R_{u}|}).s_{bit} }{E_{comm}}
\end{split}
 \label{eq:cee}
 \end{align}
where $|R_{u}|$ shows the number of requests of user $u$ during the operation time, and $s_{bit}$ denotes the request packet size in bit.
 
On the other hand, drones' mechanical energy consumption during hovering and moving depends on parameters like the drone height, the drone speed and etc. Franco et al. \cite{7101619} have done many experiments in order to formulate the power consumption of drones. The total mechanical energy consumption of the drone (in joule) can be expressed as
\begin{align}
 \begin{split}
E_{mech} = P_{hovering}.t_{hovering} + P_{moving}.t_{moving}
\end{split}
 \label{eq:flyeng}
 \end{align}
where $P_{hovering}$ and $P_{moving}$ denote the power consumption of the drone (in watt) during hovering and moving at a specific height, respectively. Moreover, the \textit{Mechanical Energy Efficiency (MEE)} (in bits/joule) can be defined as the ratio of the amount of total transmitted bits over the total mechanical energy consumption of the drone during the operation time. \textit{MEE} can be computed by

\begin{align}
 \begin{split}
MEE = \frac{(\sum_{u=1}^{U}{|R_{u}|}).s_{bit} }{E_{mech}}
\end{split}
 \label{eq:mee}
 \end{align}

\section{Simulation Results}\label{sec:simulation}
In this section, the performance of the mobile drone BS is compared with the fixed drone BS hovering over the center of target area, which will be referred to as the fixed drone BS strategy hereafter. The fixed drone BS allocates the bandwidth to the users similar to its corresponding mobile drone BS. To make the definition clearer, we should discuss about six different strategies, i.e., 
\begin{enumerate}
\item Mobile - Equal Bandwidth Division 
\item Fixed - Equal Bandwidth Division 
\item Mobile - Nearest User First
\item Fixed - Nearest User First
\item Mobile - Least Buffer First
\item Fixed - Least Buffer First. 
\end{enumerate}

Fair comparison has been conducted for the pairs of strategies (1) and (2), (3) and (4), (5) and (6), because the same bandwidth allocation scheme is used for each pair of strategies. Both mobile drone BSs and the fixed ones have the same configuration such as height, speed and transmission power.

All results are based on the simulations using MATLAB and have been averaged over 200 runs to smooth out the randomness. Users are moving according to the RWP model with the speed range of [0.2-4] m/s. The angle step is set to $\pi/36$ ($5^{\circ}$) to both reduce the complexity and ensure the reasonable accuracy of the results. Unless otherwise stated, parameter values used in simulations are according to Table \ref{tbl:values}.

\begin{table}[!t]
\caption{Values and Definition of Parameters}
\label{tbl:values}
\begin{tabular}{lll}
\hline
{\bf Parameter}              & {\bf Definition}    		  & {\bf Value}  \\ \hline
$w$					& Area Width			 &	80 m \\
$U$								&Number of Users			&5\\
$B$					& Total Bandwidth							& 10 MHz \\
$f$					&Working Frequency							& 2 GHz \\
$h$		& Drone Height 			& 10 m \\
$v$		& Drone Speed				& 20 m/s \\
$P_{tx}$		& Drone Transmission Power 						& 24 dBm \\
$\lambda$		& Mean Reading Time						& 5 sec \\
$A$		&Reference Distance Path Loss (LoS/NLoS)			 &	41.1/33 \cite{3gpp36828}\\
$\gamma$		&  Path Loss Exponent (LoS/NLoS)				 &	2.09/3.75 \cite{3gpp36828}\\
$\delta_{ue}$               & UE Noise Figure                     &  9 dB \\
$\alpha, \beta$               & Environmental Parameter for Urban Area         & 11.95 , 0.136 \\
$T$		&Simulation Time		&300 s\\
$s$		&Request Packet Size			&2 MBytes \\
$\Delta g$   &Angle Step  &$\pi/36$\\
$\Delta t$   &Time Slot  & 0.1 s\\
$P_{hovering}$   &Drone Hovering Power at $h$=10 m  &110 watt \cite{7101619}\\
$P_{moving}$   &Drone Moving Power at $h$=10 m $v$=20 m/s   &170 watt \cite{7101619}\\
$P_{moving}$   &Drone Moving Power at  $h$=10 m $v$=10 m/s &110 watt \cite{7101619}\\

\hline
\end{tabular}
\end{table}

\subsection{Simulation Time Configuration}
\begin{figure}[!t]
\centering
  \includegraphics[scale=0.45]{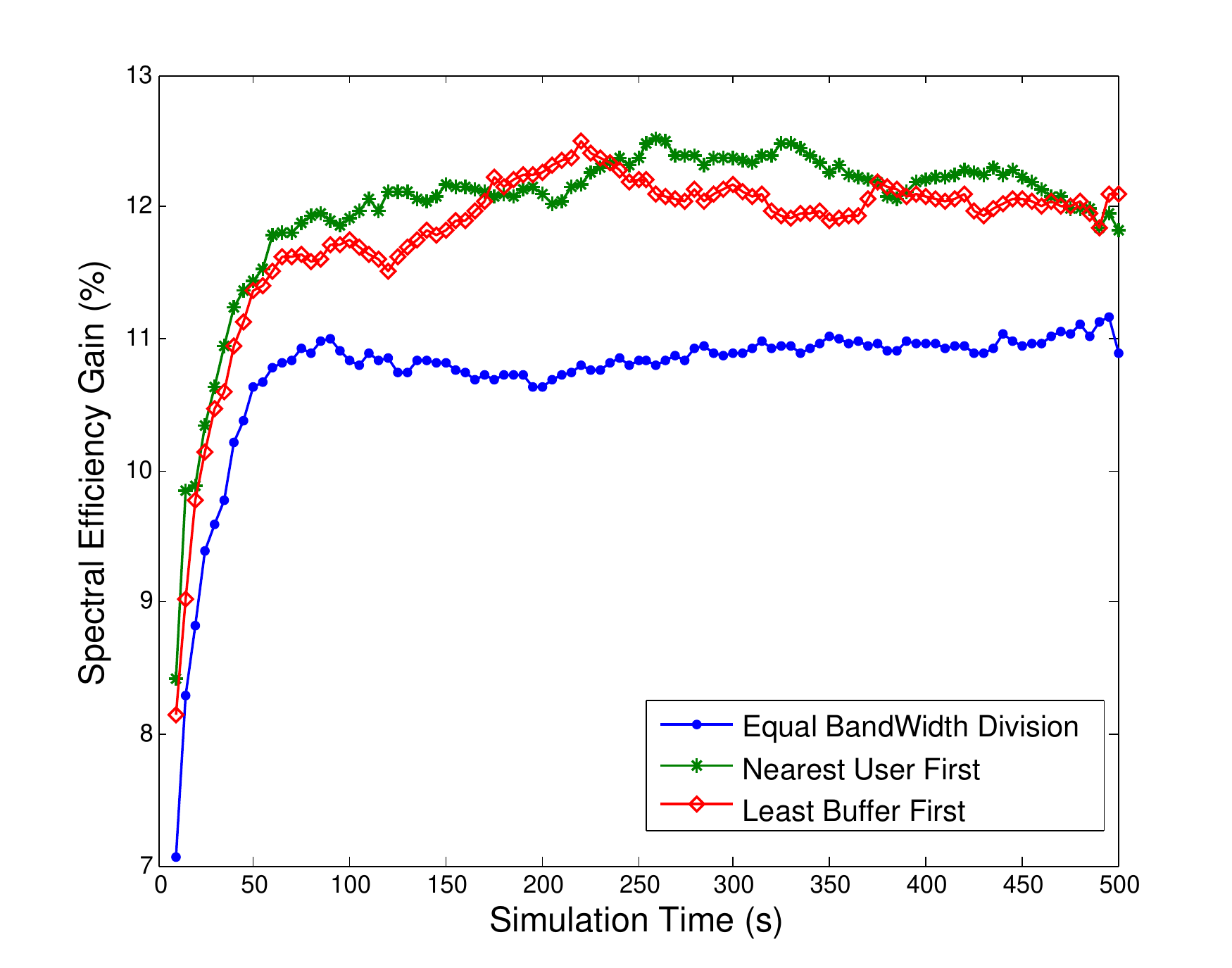}
\caption{Spectral efficiency gain of the moving drone with the speed of 20 m/s vs. the simulation time}
\label{fig:simultime20}
\end{figure}

In this section, we first analyze the (\textit{SEG}) of our proposed strategies with the assumptions of different simulation times to see how the results converge. Figure \ref{fig:simultime20} shows that the moving drone BS with the speed of 20 m/s can achieve around 13\% spectral efficiency gain. Given this figure, the system performance becomes stable after 100 sec. As a result, hereafter the simulation time have been set to 300 sec to obtain meaningful results. The results of the first 100 sec are discarded because they just warm up the simulator. 

Another important finding is that the Least Buffer First and the Nearest User First strategies give very similar performance higher than the performance of the Equal Bandwidth Division strategy. This result can be explained by the fact that dividing the bandwidth equally between all active users reduces the average spectral efficiency.

\subsection{The Performance Impact of the Area Width}
\begin{figure}[!t]
\centering
  \includegraphics[scale=0.45]{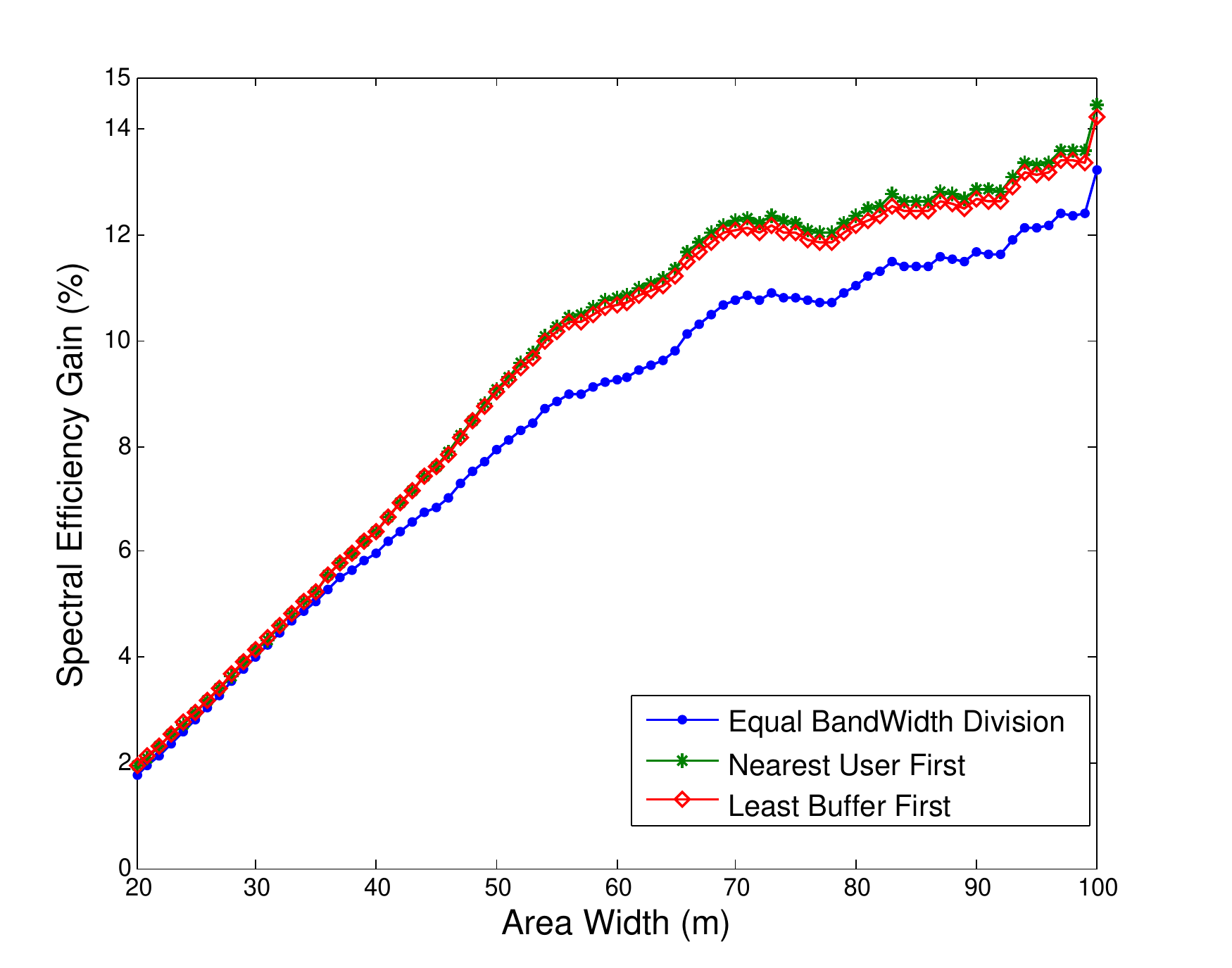}
\caption{Spectral efficiency gain vs. the area width }
\label{fig:areagain}
\end{figure}

Figure \ref{fig:areagain} shows the \textit{SEG} with respect to the area width. Similar to the previous figure, Figure \ref{fig:areagain} indicates that the Nearest User First strategy can obtain the highest gain. One important finding is that the performance gain gets larger as the area size is increasing. This is because that a small area restricts the drone movement. As a result, there is not much difference between the fixed drone BS and the mobile one. In a $100\ m\times 100\ m$ area, up to 15\% gain can be achieved.

\begin{figure}[!t]
\centering
  \includegraphics[scale=0.45]{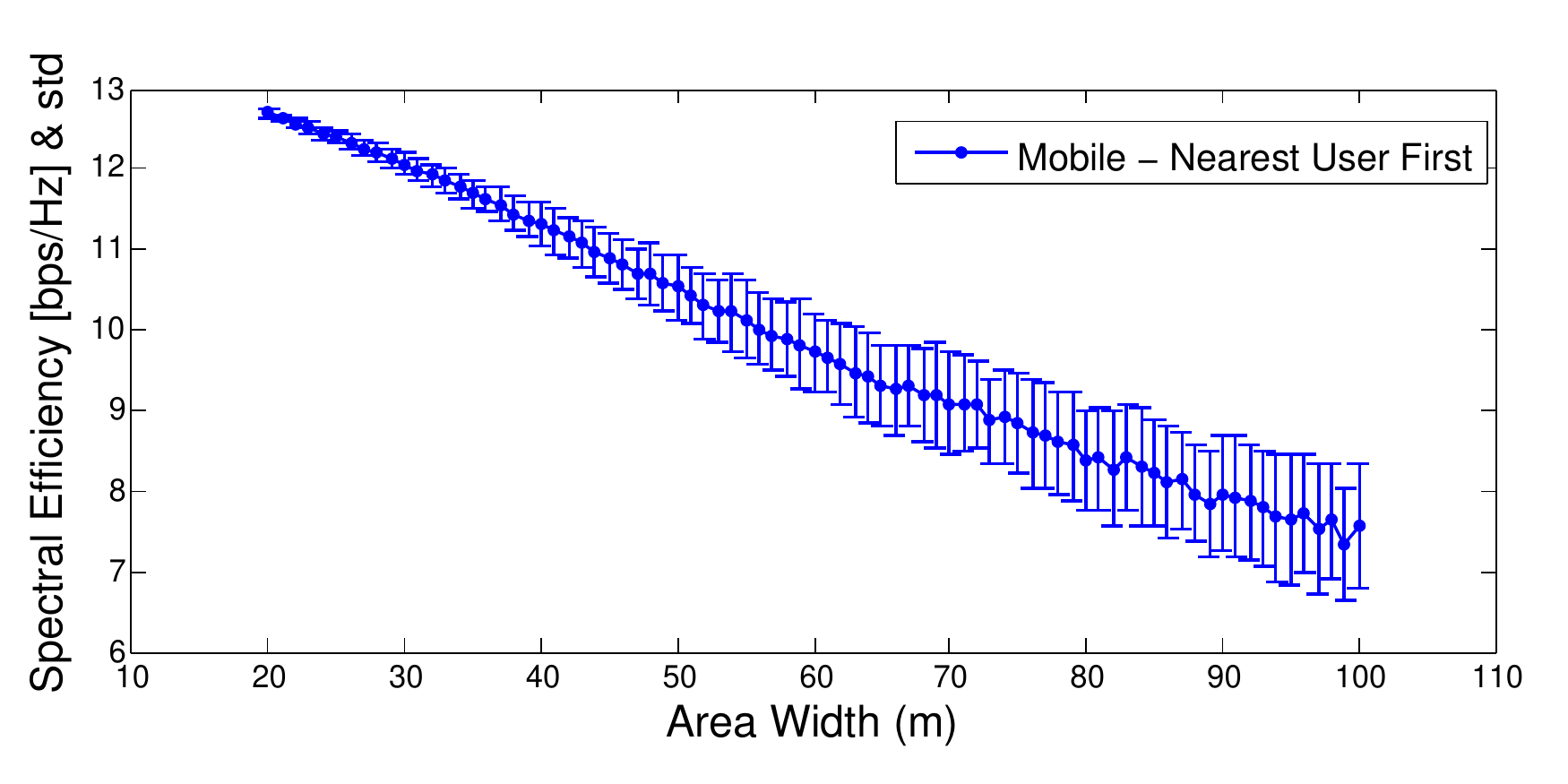}
\caption{Spectral efficiency and the standard deviation for Mobile - Nearest User First strategy}
\label{fig:areastdm}
\end{figure}

As stated before, all results are averaged over 200 runs. To check the variance of the results, the standard deviation for the value of spectral efficiency are presented as well. 
Figure \ref{fig:areastdm} illustrates the standard deviation of spectral efficiency value based on the area width for the mobile drone of the Nearest User First strategy moving with the speed of 20 m/s. The standard deviation of all simulation results is about 0.5 bps/Hz. It is interesting to note that the spectral efficiency of the system model is decreasing when the area is getting wider. It can be due to increasing the average distance between the drone and the users, and consequently decreasing the value of average spectral efficiency.

\subsection{The Performance Impact of the Drone Height }

Figure \ref{fig:heights20gain} shows the \textit{SEG} with respect to the drone height. With a higher height, the movement of the drone has a smaller effect in reducing the distance between the drone and users compared to a lower height. Therefore, the spectral efficiency gain decreases as the drone height increases.

\subsection{The Performance Impact of the User Density}
Figure \ref{fig:usergain} shows the spectral efficiency gain with respect to the user density (the number of users per $100\ m^2$ area). As can be seen from figure \ref{fig:usergain}, when the area gets denser performance gain wanes, however in more dense area the mobile drone can reach 6\% \textit{SEG}. This result can be explained by the fact that having a larger number of users means more requests distributed over the area. Thus when the drone moves towards one request, the probability of getting far from other requests increases, making the spectral efficiency suffer.
\subsection{Analysis of the Energy Efficiency}

Given the numerical results of experiments in \cite{7101619}, the power consumption of the drone moving at a specific height, is increasing as its moving speed is increasing. Because of the importance of energy consumption for drones, we first study the effect of drone speed on \textit{SEG}.

\begin{figure*}
    \centering
    \begin{subfigure}[b]{0.32\textwidth}
        \includegraphics[width=1.1\textwidth,height=1\textwidth]{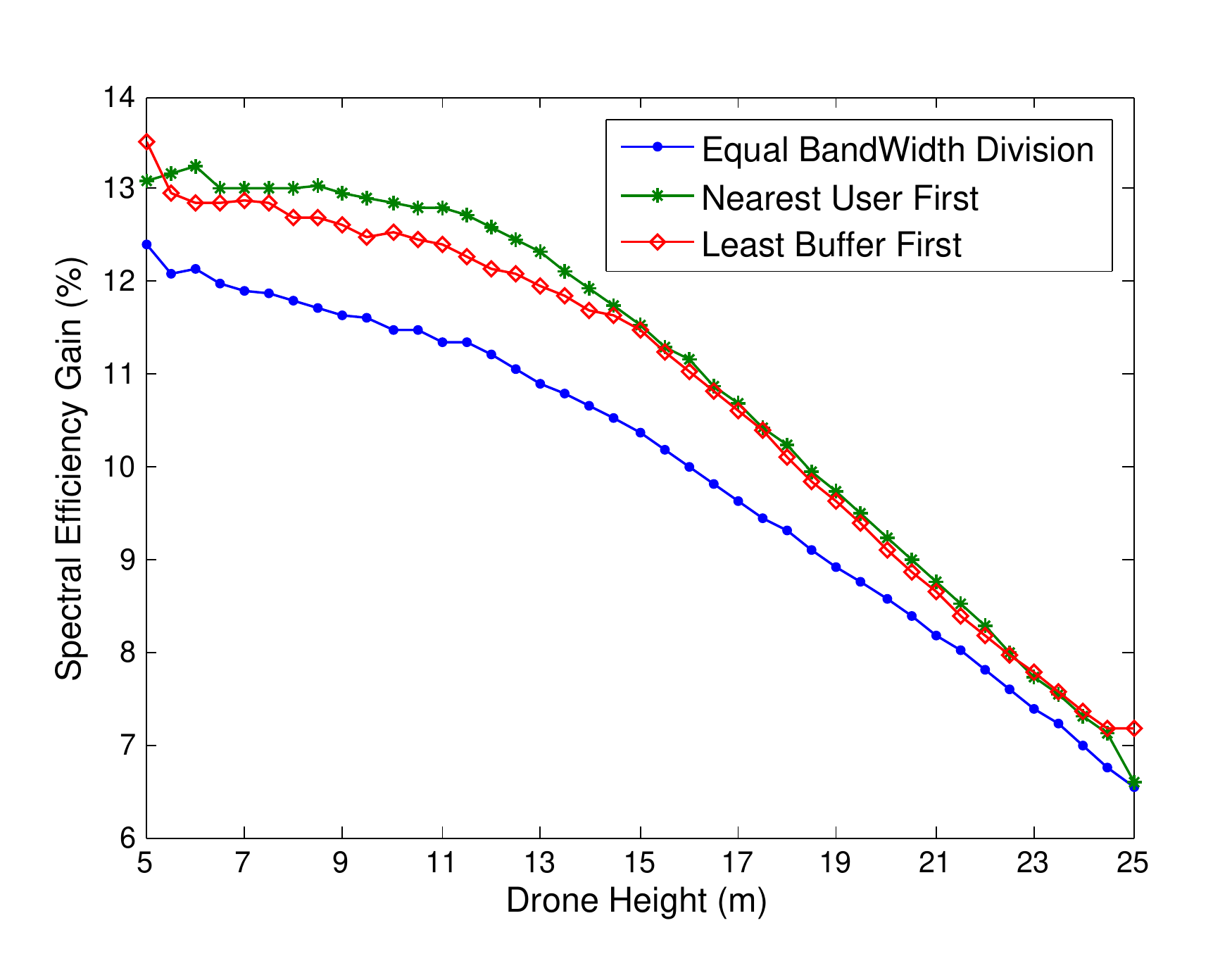}
        \caption{Spectral efficiency gain vs. the drone height}
        \label{fig:heights20gain}
    \end{subfigure}
    ~ 
    \begin{subfigure}[b]{0.32\textwidth}
        \includegraphics[width=1.1\textwidth,height=1\textwidth]{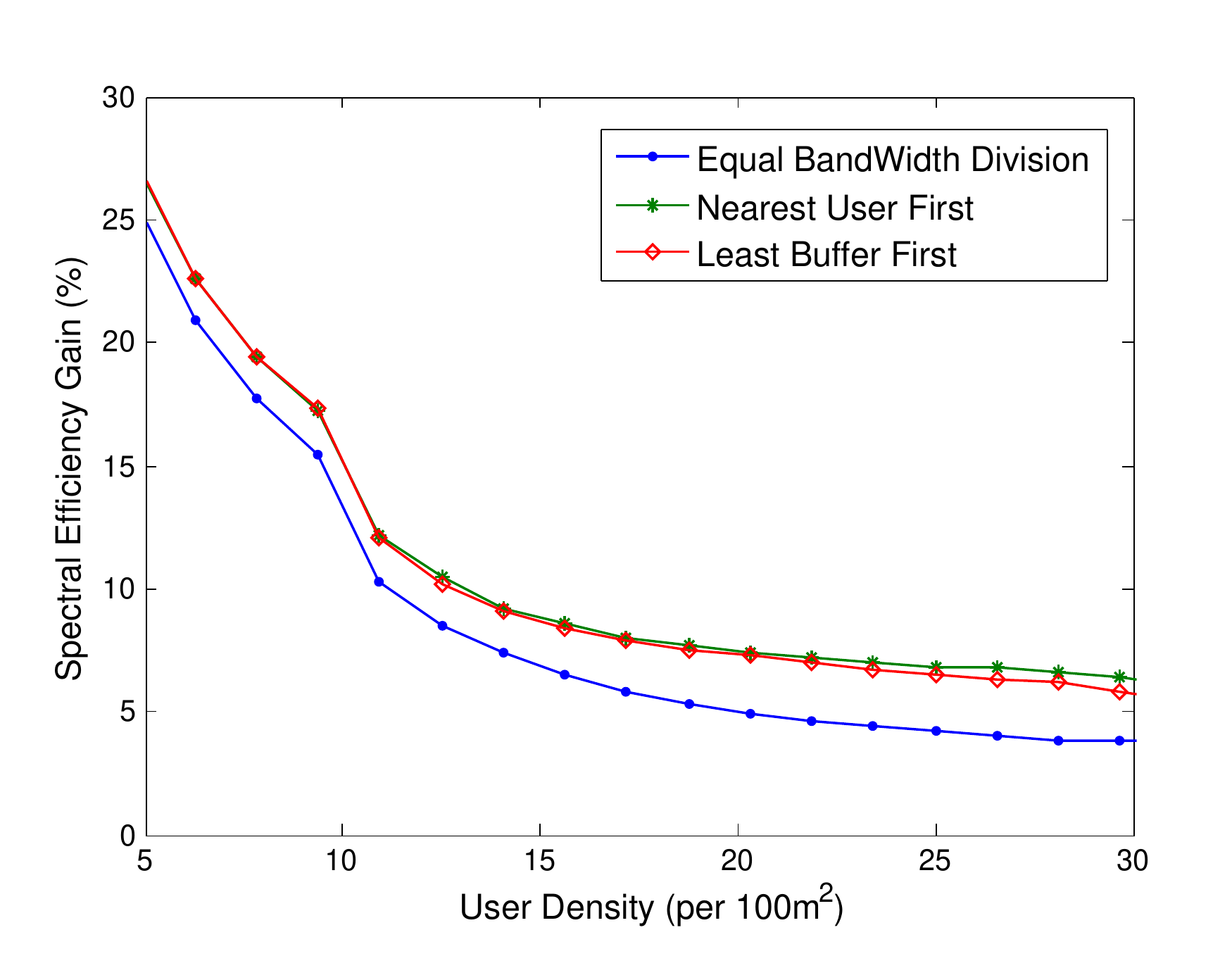}
        \caption{Spectral efficiency gain vs. the user density}
        \label{fig:usergain}
    \end{subfigure}
        ~ 
        \begin{subfigure}[b]{0.32\textwidth}
        \includegraphics[width=1.1\textwidth,height=1\textwidth]{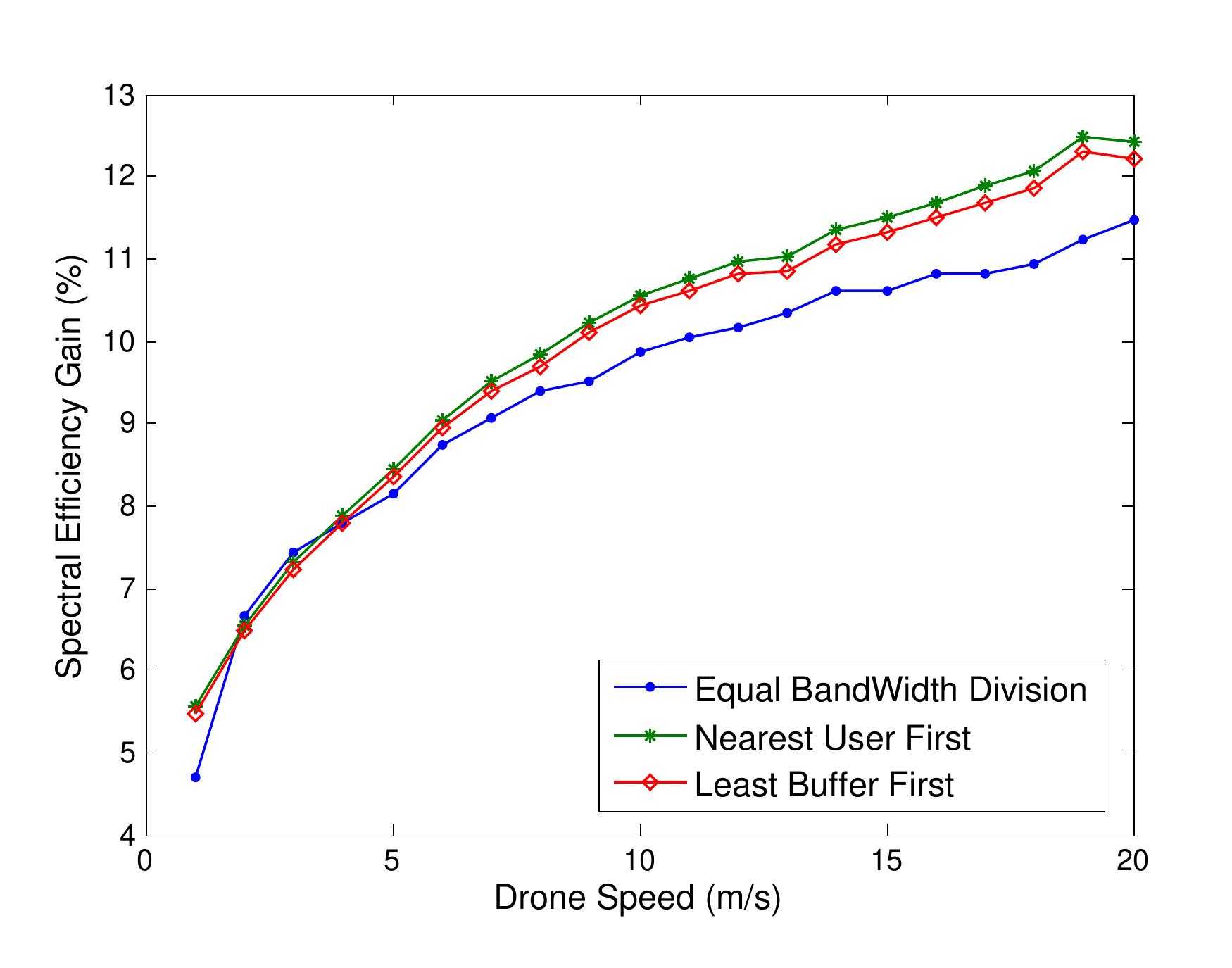}
         \caption{Spectral efficiency gain vs. the drone speed }
        \label{fig:speed}
    \end{subfigure}

    \caption{Spectral efficiency gain vs. (a) the drone height, (b) the user density, and (c) the drone speed}\label{fig:se}
\end{figure*}

Figure \ref{fig:speed} illustrates that when the speed of the drone is increasing, the gain is increasing as well. It shows a strong relationship between the speed of the drone BS and the achievable spectral efficiency gain. With higher speed, the drone BS is able to become closer to the users in shorter time and transmit data to them with higher data rates. However there is a trade-off between energy consumption and spectral efficiency gain. The power consumption of moving with lower speed, less than 10 m/s, is almost the same as the power consumption of hovering at the same height \cite{7101619}. On the other hand, moving with the speed of 10 m/s can obtain decent \textit{SEG} (10.5\% gain). Therefore, we compare the energy efficiency of the mobile drone with two different speeds, i.e., 20 m/s and 10 m/s.

\begin{figure}[!t]
\centering
  \includegraphics[scale=0.45]{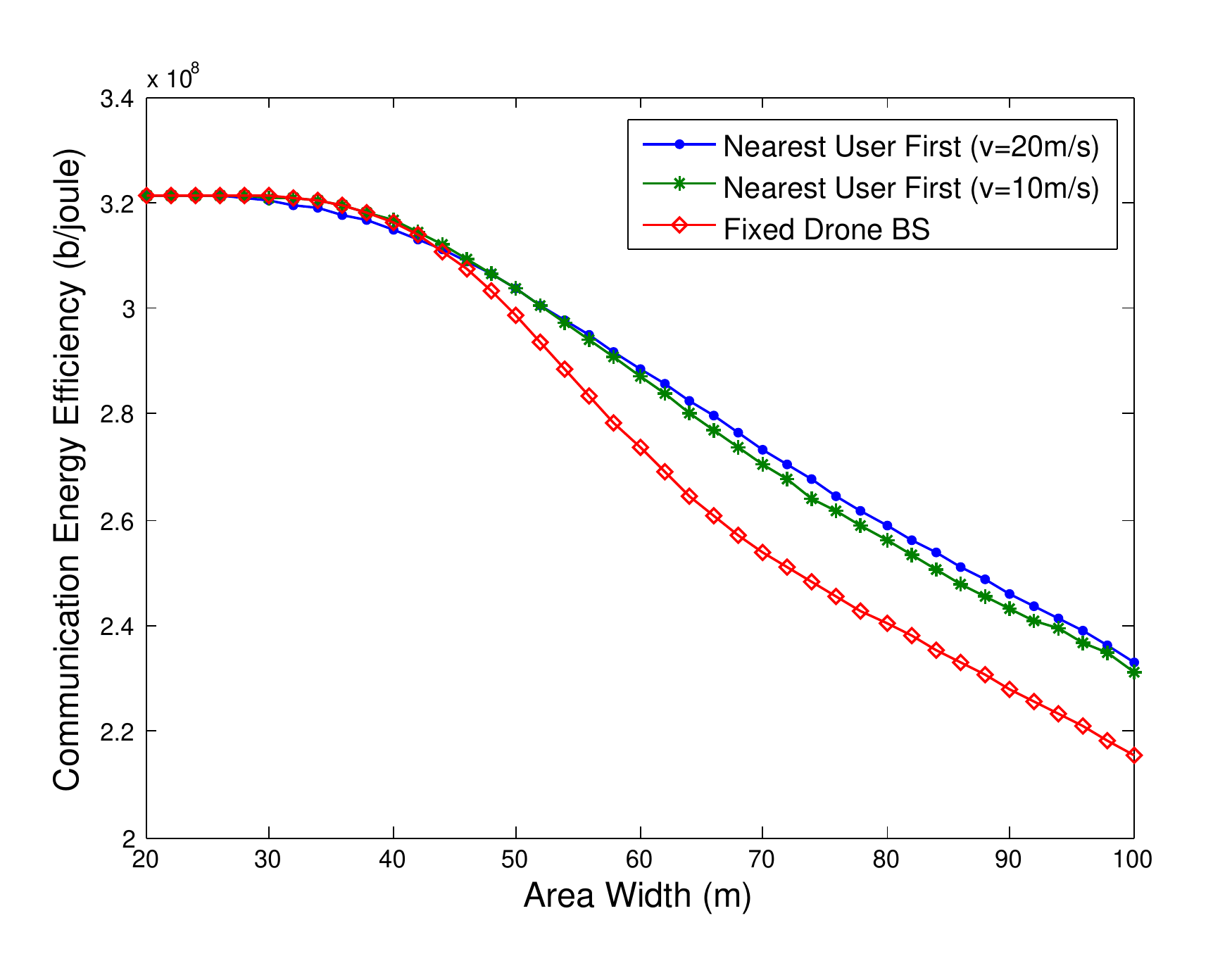}
\caption{CEE of the Fixed and the Mobile Nearest User First strategy}
\label{fig:com_eng_eff}
\end{figure}

According to the definition of the \textit{CEE} in section \ref{sec:engmodel}, figure \ref{fig:com_eng_eff} provides the CEE of the mobile drone BS for the Nearest User First strategy compared with the fixed drone BS. From this figure, the CEE of the mobile drone with the speed of 20 m/s and the speed of 10 m/s are very close to each other, and they are much more efficient than the fixed one. This is due to the reduction of communication time because of the increase of spectral efficiency value for the mobile drone BSs. Another interesting finding is that the energy efficiency decreases when the area becomes larger.

\begin{figure}[!t]
\centering
  \includegraphics[scale=0.45]{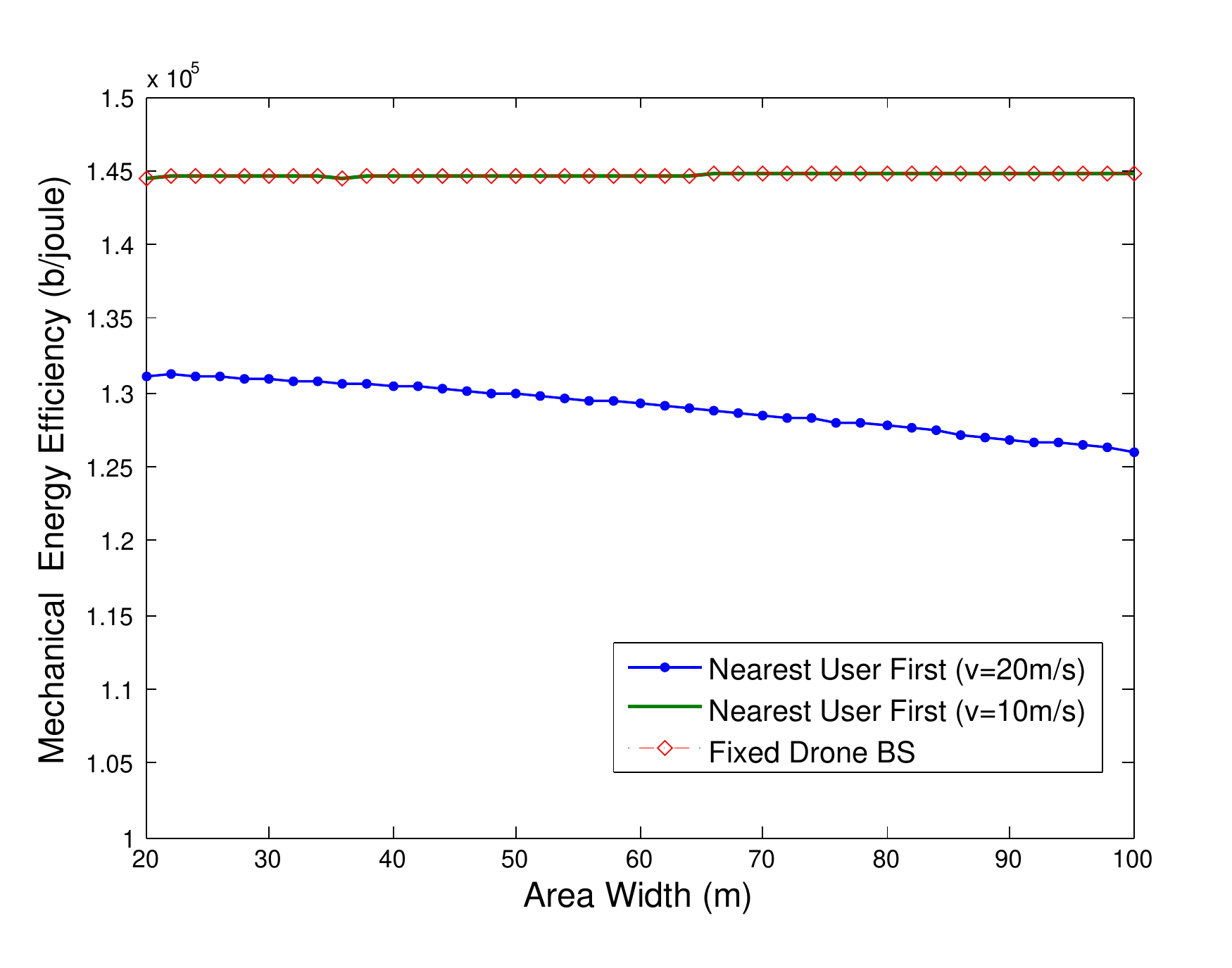}
\caption{MEE of the Fixed and the Mobile Nearest User First strategy}
\label{fig:fly_eng_eff}
\end{figure}

Moreover, the \textit{MEE} of our proposed strategies is analyzed.
As shown in figure \ref{fig:fly_eng_eff}, the fixed drone BS is much more energy efficient than the mobile drone with the speed of 20 m/s. Because of the similar value of hovering power consumption and moving with the speed of 10 m/s, the mobile drone BS with the speed of 10 m/s has the same MEE as the fixed drone BS. This result is very encouraging which implies that we can achieve 10.5\% spectral efficiency gain and higher \textit{CEE} by using mobile drone BSs with the speed of 10 m/s compared with the fixed one, while retaining a similar \textit{MEE}.

\subsection{Summery on the Results}
To sum up, the proposed algorithms show considerable gains in terms of spectral efficiency compared with the fixed drone BS scheme. By reducing the speed of the drone, the spectral efficiency gain reduces however better results for energy efficiency can be obtained. It is also found that \textit{SEG} improves by reducing the height of the drone and increasing the area size. Our proposed methods can even obtain acceptable \textit{SEG} in dense area. Among three different proposed strategies, the Nearest First User strategy outperforms other ones.

\section{Conclusion and Future Work} \label{sec:conclusion}
We have proposed dynamic repositioning of the BS as a novel method to increase spectral efficiency of drone small cells. We proposed three simple algorithms that can be used to autonomously reposition the drone in response to user activities and mobility. Our analysis reveals that there is a tradeoff between spectral efficiency and energy efficiency of the drone. We have shown that 10.5\% increase in spectral efficiency is possible without any negative impact on the energy efficiency of the drone. Considering multiple drones and 3D movement would be our future work. Besides, we will consider stochastic geometry analysis in our future work.

\bibliographystyle{IEEEtran}
\bibliography{gc_uav_archive}

\end{document}